\documentclass[11pt, oneside]{article}   	% use "amsart" instead of "article" for AMSLaTeX format
\usepackage{geometry}                		% See geometry.pdf to learn the layout options. There are lots.
\usepackage{graphicx}				% Use pdf, png, jpg, or eps§ with pdflatex; use eps in DVI mode
\usepackage{amssymb,hyperref}
\usepackage[utf8]{inputenc}

\title{`Passion for Earth': A New Beginning}
\author{Francesco Vissani\\ \normalsize INFN, Laboratori Nazionali del Gran Sasso}
\date{}							% Activate to display a given date or no date

\begin{document}
\maketitle
\begin{abstract}
In this essay, I discuss an interdisciplinary science that is just blossoming: that of geo-neutrinos. I begin with a couple of episodes from the history of thought, showing the deep roots of Earth science and its many connections with microphysics. I then recall the stage of full maturity reached in the knowledge of neutrinos, which allows one to argue the full reliability of recent measurements. I conclude with a discussion of the state of the field and its prospects for the near future, and with a dedication to the memory of a colleague and friend, Giovanni Fiorentini, a pioneer of these studies.\end{abstract}
%\section{}
%\subsection{}

%\LARGE
\parskip0.95ex
\section{Introduction}
The discussion at the workshop {\em``Passion for the Earth: A New Era for Geoscience?''}  was thorough and comprehensive; see~\cite{puw} for the webpage and 
Appendix~\ref{summ} for a brief summary.  My task was to wrap up the discussion; but before commenting on the reported progress--primarily, the high-quality geo-neutrino observations--I suggest we take a moment to assess the big picture. I propose to do so by reviewing some crucial steps that have been taken that have led to important insights about the Earth.

The intention is to speak lightly, recalling stories rather than describing history in detail. I will do so, however, with certain goals in mind: first, I will emphasise how certain ideas have had lasting and sometimes  unexpected effects; then, I will try to argue that links between different disciplines are valuable, if not necessary, for renewing and revitalising the various fields of research (which also applies to Earth sciences and particle physics). I will dwell on moments of change, recounting not only the results but also the obstacles encountered in the progress of knowledge - particularly those attributable to the persistence of originally fruitful ideas, which at some point crystallise into unverified opinions that can and must be tested by observations.

The structure of this contribution is the following: In section~\ref{sez1} we present some ancient ideas about the Earth, showing their close connection with those  concerning the nature of matter. In section~\ref{sez2} we focus on the time of  transition to current views, underlining that they are quite recent. In section~\ref{sez3} we talk about how neutrino science was born, and how their interactions and curious properties have been understood. Section~\ref{sez4} (and the appendix) regards the topic of interest for this workshop, geo-neutrinos, including an evaluation of its pace of progress and current status. Finally, in section~\ref{sez5} we present a summary and outline the perspectives.

%
%I  will discuss: 1) certain connections of Earth science; 
%2)~the reliability and relevance of neutrino studies;  
%3)~the birth of a new discipline, geo-neutrino science. Then I conclude with a discussion and a   
%dedication to the memory of G.~Fiorentini.

\section{An old passion but not exclusive} \label{sez1}
We are by nature travelers, and the passion for reasoning about the Earth on which we live is as old as humankind. 
This passion was slowly transformed into coherent reflections, taking first the form of philosophy and then that of science, as far back as ancient Greek times.
E.g., Theophrastus attributes the discovery of the shape of the Earth to Parmenides, as reminded by 
Diogenes Laertius (although the same tells us that Zeno attributes it to Hesiod, and others to Pythagoras). 
See the first part of Table~\ref{tab1} for a list of some important steps forward in the reasoning on these issues.

\begin{table}
\centerline{
\begin{tabular}{l|c|c}
{\bf Parmenides}	& 525?-\,450 BC	& Spherical Earth  \\
{\bf Anaxagoras} & 	496 - 428 BC &	Constitution of Celestial Bodies \\
{\bf Aristarchus}	& 310 - 230 BC	& Heliocentrism \\
{\bf Eratosthenes}	& 267 - 194 BC	 & Accurate Earth Measurement \\
{\bf Hipparcus}	& 190 - 120 BC	& Trigonometry, Astronomy \\ \hline\hline
{\bf Empedocles}	&V century BC	& Plurality of Elements \\
{\bf Leukippus,  Democritus} &	V  century BC	& Atoms (Arch\'e) \\
{\bf Euclid}	&IV century BC	& Optics \\
{\bf Ctesibius}	&III  century BC	 & Pneumatica \\
{\bf  Archimedes}	&287 - 212 BC	& Hydrostatics, Mechanics \\
%Philo &	280 - 220 BC	& Automata
\end{tabular}}
\caption{\em\small 1st column, ancient Greek thinkers; 2nd one, approximative dates of their life; 3rd one, some important topics they examined. Top: Themes related to the Earth and the Cosmos. Bottom: Topics relating to the constitution of matter, its knowledge and its use.\label{tab1}}
\end{table}

A very important figure in this discussion is Anaxagoras, who suggested that celestial bodies are made of the same matter that surrounds us, such as rocks and stones.
According to this philosopher, the Sun's light is due to the fact that it is composed of incandescent matter, such as that we see in the lavas of volcanoes or in the melting of metals.  It is not entirely useless to recall that Anaxagoras was tried and condemned for these ideas.\footnote{We are reminded of this in Plato's book {\em The Laws}, in which punishment is invoked for those who degrade the idea of the gods, which every Greek could see in the sky with his own eyes.}

The second part of the table, as interesting as the first, points instead to thinkers engaged in reflection on the intimate nature of matter. 
These include the other pluralist philosophers: Empedocles, who supported the idea of four fundamental elements, corresponding to aspects of our 
everyday experience; and also the two celebrated atomists, Leucippus and Democritus, who proposed the idea that matter consists, at a level not directly perceptible, of 
indivisible atoms and vacuum.  Table~\ref{tab1} then lists the names of some of the major mathematicians of Hellenism, who testify to the evolution of Greek thought 
toward accomplished mathematical formalizations. For an articulate defense of this point of view, see~\cite{russo}.

The two strands of enquiry - that of larger things and that of smaller things - proceed by extension of everyday experience. It is very interesting to note that they have intertwined and interpenetrated since antiquity, particularly in the reflections on the Earth. A notable example is provided by Erwin Schr\"odinger's essays on the Greeks \cite{erwin}; here is an evocative commentary on the famous dogma of the first philosopher and mathematician, Thales
\begin{quote}
{\em\small A rather interesting detail, reported by several doxographers as Thales' opinion, is that the land floats on the water `like a piece of wood'; which must mean with a considerable part of it immersed. This [...]
%recalls on the one hand the old myth about the isle of Delos wandering
%around erratically until Leto there gave birth to twins, Apollo and Artemis; but it 
is also amazingly akin to the modern theory of isostasy, according to which the continents do float on a liquid, though not exactly on the water of the oceans, but on a heavier molten substance below them.

}
\end{quote}
In the same connection, I would like to recall a theorem of Archimedes concerning the spherical shape of celestial bodies, 
the proof of which is based on the assumption that they are liquid {\em at some stage} of their life. 
See again \cite{russo} for  useful elaborations of the point.

In the following discussion, we will see that the hypotheses of the Greek philosophers (in their original form) will continue to have a very significant influence on scientific thought, right up to the times very close to us.

\section{Transitioning to new ideas} \label{sez2} % : intertwining, obstacles and synergies}
In this section, we examine two important episodes in the transition from ancient to modern ideas,
occurred about one century ago. 
In particular, we will look at the way in which the ideas of Anaxagoras were
Anaxagoras were challenged and eventually superseded; and also 
how progress in the understanding of our planet and celestial bodies could only be made after abandoning the idea of an indestructible atom, central to the thought of the atomists.

\begin{table}[t]
\centerline{
\begin{tabular}{c|l} 
\hline\\[-1ex]
& {\large\sc\hskip4cm HYDROGEN} \\[1ex] \hline
1766  & {\bf Cavendish} identifies it as an element; burning it, water is produced (1781)\\
1783  & {\bf Lavoisier \& Laplace} propose to call it water generator ({\em hudro-gen$\bar{e}$s})\\
1815  & {\bf Prout} claims it is the simplest and most fundamental element\\
1913  & {\bf Bohr} uses it as a `Rosetta stone' to understand the atom\\
1925  & {\bf Cecilia Payne} shows that it is the most common substance in the Sun\\
1948  & {\bf Alpher, Bethe, Gamow} primordial nucleosynthesis starts with hydrogen\\ \hline\\[-1ex]
& {\large\sc\hskip4.65cm HELIUM} \\[1ex] \hline
1868  & {\bf Janssen} observes a new line in the solar spectrum\\
1868  & {\bf Lockyer \& Frankland} name the element after the Sun's Greek name\\
1881  & {\bf Luigi Palmieri} reveals that line in the incandescent lava of Vesuvius\\
1895 & {\bf  Ramsey} discovers helium in a uranium ore\\
1895 & {\bf  Cleve \& Langlet} measure its atomic weight\\
1907 &  {\bf Rutherford \& Royds} demonstrate that   \protect{$\alpha$}-rays are helium nuclei\\ \hline
\end{tabular}
}
\caption{\em\small Important modern episodes for the current understanding of the lightest elements and their role.
Upper part of the table: events regarding hydrogen.
Lower part of the table: events regarding helium.
Hydrogen was produced also previously  by investigators combining acids and metals, such as 
Paracelsus ($\sim$1500), van Mayerne (1620), Boyle (1671). On the contrary history of helium is entirely recent.\label{tab2}}
\end{table}

\subsection{The road to understanding the composition of the Sun}
Today, when people talk about stars (or possibly the Sun) it is taken for granted that they are made mainly 
of hydrogen and helium. What is only rarely remembered is that these intellectual acquisitions are less than a century old,
and how complex was the travail to overcome the ideas of the Ancients.  
Let us summarise some interesting events relating to the moment of transition, 
referring to Table~\ref{tab2} for a more systematic summary of the crucial discoveries about these two important chemical elements.

Let us begin with the so-called ``meteoric theory'' of Sun formation. 
This theory, which had previously been considered for planets, and in particular the Earth, required extrapolation to be considered for stars as well. 
One of its influential proponents was the founder of the famous journal Nature, Lockyer \cite{lock}.

 Another authoritative supporter  of ``meteoric theory'' was the astronomer  Henri Norris Russel, universally known for his diagram of the 
 distribution of the stars in the  temperature--magnitude plane. In 1914, he wrote
a paper \cite{russ} to argue that the spectroscopic observations of solar and terrestrial matter (when heated sufficiently) lead to quite similar outcomes, which apparently 
supported  the idea, that the composition of the two bodies was similar.  
He stated that, ``non-metallic'' elements
were scarce in the Sun; furthermore supporting this opinion  speculating on a link with the recently discovered radioactive phenomena. In short, he suggested that the Sun is made up  of ``metallic'' (heavy) elements just 
as the Earth; note that this corresponds in a narrow sense to the ideas of Anaxagoras.\footnote{A modern reader might want to counter: 
{\em The simplest elements - hydrogen, helium! - are the most common in the universe, they cannot but play an essential role in stars}.
One might be tempted to support this thesis by relying on Prout's views, as in Tab.~\ref{tab2}. But (assessing the views in their historical context) Prout was not considered authoritative in Russell's time, a transfer of ideas from atomic physics to astrophysics would have taken time, and the big-bang theory was far from arriving.}

The first scientific work to show that the Sun is composed of hydrogen and helium is 
Cecilia Payne's 1925 doctoral thesis \cite{ceci}. In this study, it is observed that the surface temperature of the Sun 
is relatively low, so that electrons of these two atomic species are hardly excited; once this is taken into account,  
quantitative deductions can be made. However, the supervisor of her doctoral thesis was Russell himself, who objected to this important 
result being explicitly put in written. Fortunately Payne included the table of the results, which four years later were confirmed by an investigation of 
Russell himself \cite{russ2}. See also:
historian of science DeVorkin's work on the affair~\cite{devo} and Moore's fine book on the figure of Cecilia Payne~\cite{moore}.\footnote{The first is a very informed research work with methodological aims, such as commending cautious attitudes in scientific investigations, or the exercise of `political acumen'. I do not judge it, but I cannot help feeling that some facts are interpreted in a somewhat conciliatory way. Indeed, Russell's position in 1914, based on grossly incomplete data modelling, is not considered a moral obligation, but is justified as being `standard picture'; at the same time, his authority in maintaining reservations about Saha's theory in 1925 is taken for granted. The second paper is a biography; it specifically describes the difficulties Cecilia Payne encountered with her doctoral thesis.}

\subsection{One hidden hypothesis concerning microphysics}

We would now like to highlight a second hypothesis,  also of Greek roots, 
which
has continued to guide/bias scientific thought for a long time, without being perceived as such.
It is basically the idea that at a fundamental level matter is to some extent compressible, but essentially immutable. 

We can see its effects in various instances.\\
$\star$ For example, Laplace's speculated on the existence of  
stars so large, that light cannot emerge - a kind of germ of the idea of the black hole, from 1796, see \cite{lap}.
In practice, he proposed a star 250 times larger than the Sun but only 4 times denser; thus the escape velocity 
is equal to $c$. A question arises, why not increase the density instead - which would resemble more the current picture of black holes?  \\
$\star$ A similar question arises considering the very famous diatribe \cite{debate} 
that pitted William Thomson  against Charles Darwin, concerning the estimation of the age of the Earth. 
As is well known, Thomson (not yet Lord Kelvin at the time) used a model of the Earth to estimate its age,\footnote{Today we know that the model was incorrect and incomplete: constant thermal conductivity was assumed  and radioactivity was unknown.}
finding a result too short, but that 
was supported by his previous estimate of the Sun's age. The latter was obtained from the relationship between the gravitational energy that the Sun possesses and 
the power it emits (the luminosity) which is measured.
The gravitational energy was estimated by the formula $G_N \times M_\odot^2/R_\odot$, where $G_N$ is Newton's constant, 
and $M_\odot$ and $R_\odot$ are the mass and radius of the Sun. Again we may ask: and why not imagine that in the centre of the Sun the density can grow much greater than the average one, thus increasing the estimated potential energy available?

In both cases, one perceives a kind of repulsion to the idea of believing in the possibility of matter behaving differently from what is known, even in decidedly extreme and never before explored situations,
such as the center of the stars.

One spontaneous explanation for this reluctance is suggested by everyday experience: under normal conditions, no material much denser than gold can be found on Earth. In this argument we can perceive echoes of the views of Empedocles and Anaxagoras. 

Another explanation, more radical but perhaps more profound, draws on the ideas of Democritus: at most we can eliminate the vacuum between atoms, but, by their very nature, they are immutable. There is a maximum density, insuperable in principle \cite{enriques}.

To understand that this is the case, let us read a few sentences from an article published in the first volume of Nature (1870), again by William Thomson 
\cite{wt}:
\begin{quote}
{\em\small The idea of an atom has been so constantly associated with incredible assumptions of infinite strength, absolute rigidity, mystical actions at a distance, and indivisibility, that chemists and many other reasonable naturalists of modern times, losing all patience with it, have dismissed it to the realms of metaphysics, and made it smaller than ``anything we can conceive''. But if atoms are inconceivably small, why are not all chemical actions infinitely swift? Chemistry is powerless to deal with this question, and many others of paramount importance, if barred, by the hardness of its fundamental assumptions, from contemplating the atom as a real portion of matter occupying a finite space, and forming a not immeasurably small constituent of any palpable body.

}
\end{quote}

A few decades later, the atom would be broken and, around the same time, radioactivity would be discovered. Microphysics was about to be completely revolutionised, with great beneficial effects on our understanding of the Sun and the Earth.

\section{Almost invisible fragments of matter}\label{sez3}

In this section,   we recall the microphysics relevant to the discussion of geo-neutrinos,  
examining the cross section that interests us (Sect.~\ref{ss1}) and the phenomena to which these particles are subjected during their propagation (Sect.~\ref{ss2}). 

In the spirit of the rest of this contribution, and in the hope of favouring the transfer of information with the relevant scientific communities, 
I will emphasise the way in which ideas have been clarified and consolidated over the years, rather than focusing exclusively on the latest results.

\subsection{From neutrino discovery to their observation}\label{ss1}
The existence of a new, very light and almost invisible neutral particle was first intuited when studying radioactive phenomena.
It had been known for many years that there are transformations of certain nuclei in which an electron is emitted. Since at the time it was believed that 
the electron was enclosed in the nucleus, it was expected that it should be released with a fixed energy; {\em but that was not what was seen.}
Pauli then proposed (1930) that there was another actor involved, namely the neutrino, which is also emitted and, 
without being seen, takes away some energy - in a sense, it steals it.

Shortly after, when the existence of the neutron was recognized (1932),
the model of the nucleus was modified.
This solved various problems concerning the spin and magnetic moment of nuclei. But in doing so, one had to admit that the neutron itself had to 
behave in a very strange way, `spontaneously disintegrating' or better, spontaneously transforming over time into a proton, an electron and an antineutrino,
as summarised by the formula,
\begin{equation}\label{nd}
n\to p+ e^- + \bar\nu_e
\end{equation}
It was Enrico Fermi who proposed 
this idea and provided its mathematical description (1933) \cite{furmo}. Fermi's theory also made it possible to predict the existence of new processes, 
connected to the previous one `algebraically', so to speak: just swap particles with antiparticles from the right member to the left and vice versa, 
or revert the direction in which the transformation takes place - thus,  symbolically, change the direction of the arrow. 
In this way, the previous process is related to this one, % pointed out by Bethe and Peierls
%Among the most interesting processes related to the previous one is this one
\begin{equation}\label{ibd}
 \bar\nu_e + p \to n + e^+
\end{equation}
This is not a disintegration, 
but rather a reaction: one in which the antineutrino collides with a hydrogen nucleus (a proton $p$) and transforms into  
two different particles, a neutron $n$ and an anti-electron $e^+$. While Eq.~\ref{nd} is entirely characterised by the decay time of the neutron $\tau_n$, 
Eq.~\ref{ibd} 
is characterised by the so-called cross section $\sigma_{\bar\nu_e p}$, which is, roughly speaking, the area of the anti-neutrino as seen by the proton being impacted.\footnote{For ease of discussion, above I have used the modern notation, which  takes it for granted that neutrinos and antineutrinos are distinct and that each is associated with an electrically charged particle. Therefore, in the  formula  of neutron' disintegration I talk of $\bar\nu_e$, namely the the anti-neutrino  (as indicated by the bar above the symbol $\nu$)  associated with the electron (as reminded by the subscript $e$).
But I warn the reader interested in historical considerations that arriving at these positions took time. E.g., Fermi originally considered the decay $n \to p+e+\nu$, moreover  adopting the `Dirac sea' interpretation; the implications of crossing symmetry were not highlighted until later, thanks to Wick, Bethe \& Peierls and others.},

 \begin{figure}[t!]
\centerline{\includegraphics[width=0.73\textwidth]{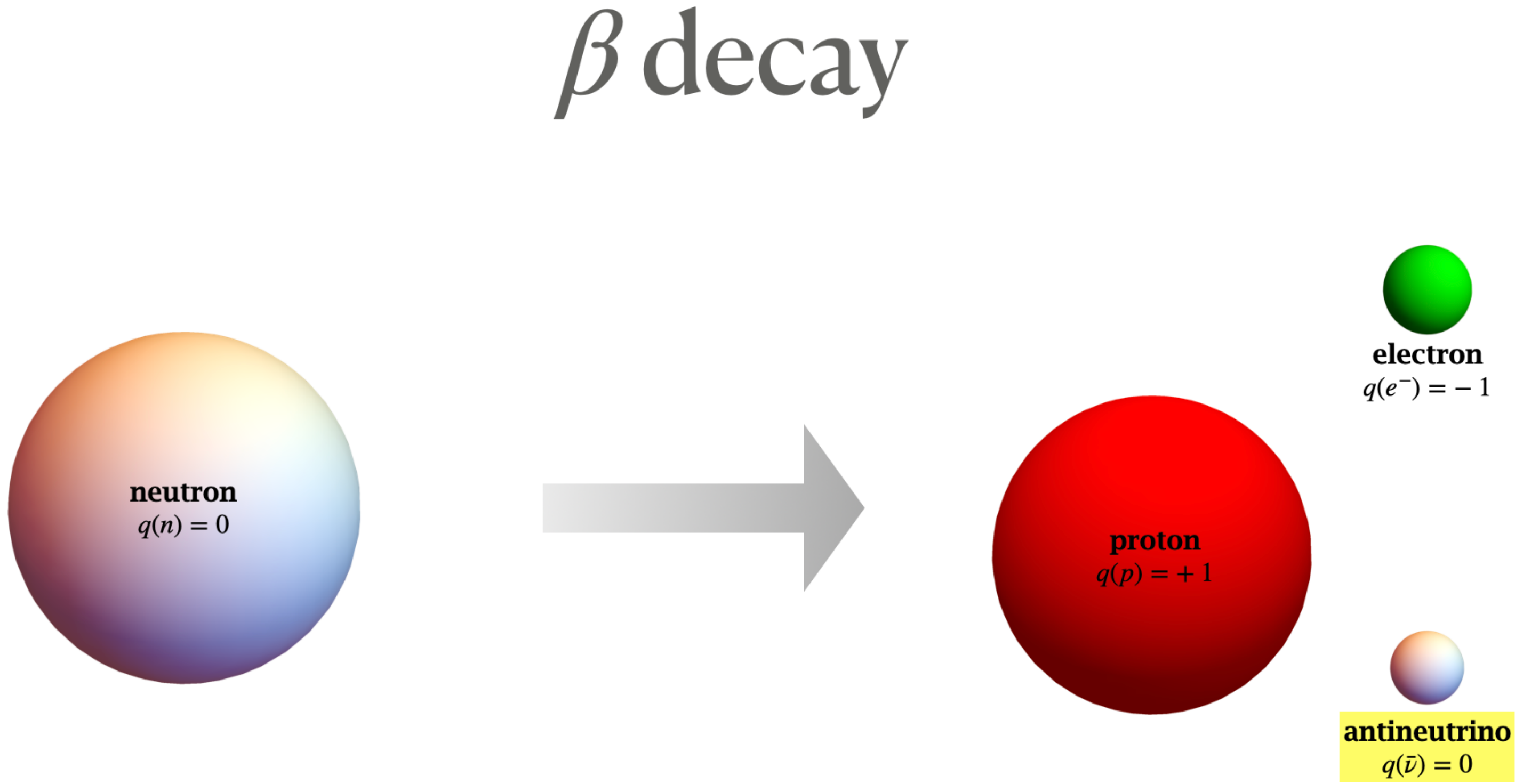}}
\vskip2mm
\centerline{\includegraphics[width=0.73\textwidth]{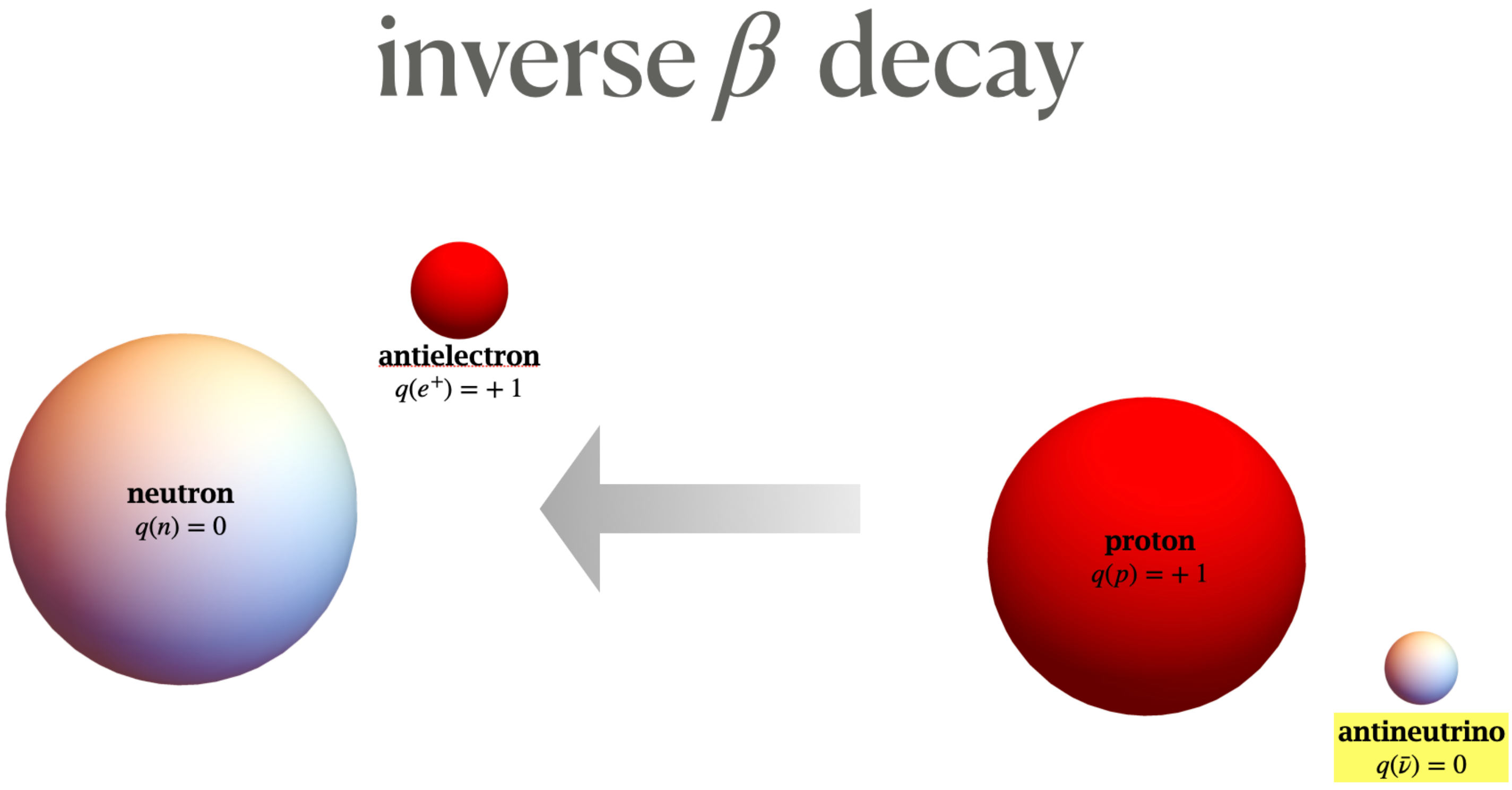}}
\caption{\small\sf\em  The graphical comparison of `$\beta$ decay' (in which a neutron transforms into a proton, an electron and an antineutrino)
and the process of `inverse $\beta$ decay' (in which the antineutrino interacts with a proton and transforms into a neutron and a positron) emphasizes their close relationship - crossing symmetry.
Indeed, in the context of relativistic field theory, adopted since Fermi's time, the two processes are described by the same Hamiltonian.}\label{fig1}
\end{figure}
In 1934, Bethe and Peierls
observed that the ratio of the two quantities $\tau_n/\sigma_{\bar\nu_e p}$  could be estimated roughly with 
simple considerations\footnote{The estimate was given by $\tau_n/\sigma_{\bar\nu_e p} \sim 
c \times (M/\hbar)^3$, where $c$ is the speed of light, $\hbar$ the reduced Planck constant and $M$ a mass characteristic of these reactions,
e.g., the mass difference between neutrons and protons, or the mass of the electron.}: thus, from the measurement of $\tau_n$ one could get an idea of the size of  
the cross section of antineutrinos interaction onto a proton, called `inverse $\beta$ decay' - see Figure~\ref{fig1}. 
Since $\sigma_{\bar\nu_e p}$ turns out to be very small, this argument 
indicated that it was very difficult to see the antineutrino. This inclined Bethe and Peierls toward a  
pessimistic attitude, concluding that  \cite{cazzo}
\begin{quote}
{\em\small there is no practically possible way of observing the neutrino.}
\end{quote}
But it was not until 20 years later, in 1956, that Reines and Cowan, 
succeeded in revealing the effects of this reaction. This was achieved by using detectors containing large masses of protons and exposed to copious streams of antineutrinos, which were emitted by the first nuclear reactors ever built, through reactions of the same type as the one in Eq.~\ref{nd}.
The existence of antineutrinos had been conclusively demonstrated, as eventually recognised by the 1995 Nobel Prize in Physics awarded to Reines.

In short, the concept {\em neutrino} emerged from a wrong theory of the atomic nucleus, the hypothesis that it existed (or could be seen) was fraught with doubt, the first observations were very imprecise and even Fermi's theory was revised in some important details. Nevertheless, the idea has been fully consolidated over the years and today particle physicists are completely comfortable with what started out as an 'invisible particle'.

It should be noted in passing  that the signal of Eq.~\ref{ibd} 
is well characterised in a scintillating detector, as it leads to two distinct observable manifestations: the release of the positron energy 
(from scintillation and annihilation), and the signal from deuteron formation: $n+p\to D+\gamma(2.2\mbox{ MeV})$, a process that occurs with an observable delay.

\subsection{Spontaneous neutrino transformations}\label{ss2}

\begin{figure}[t!]
\centerline{\includegraphics[width=0.82\textwidth]{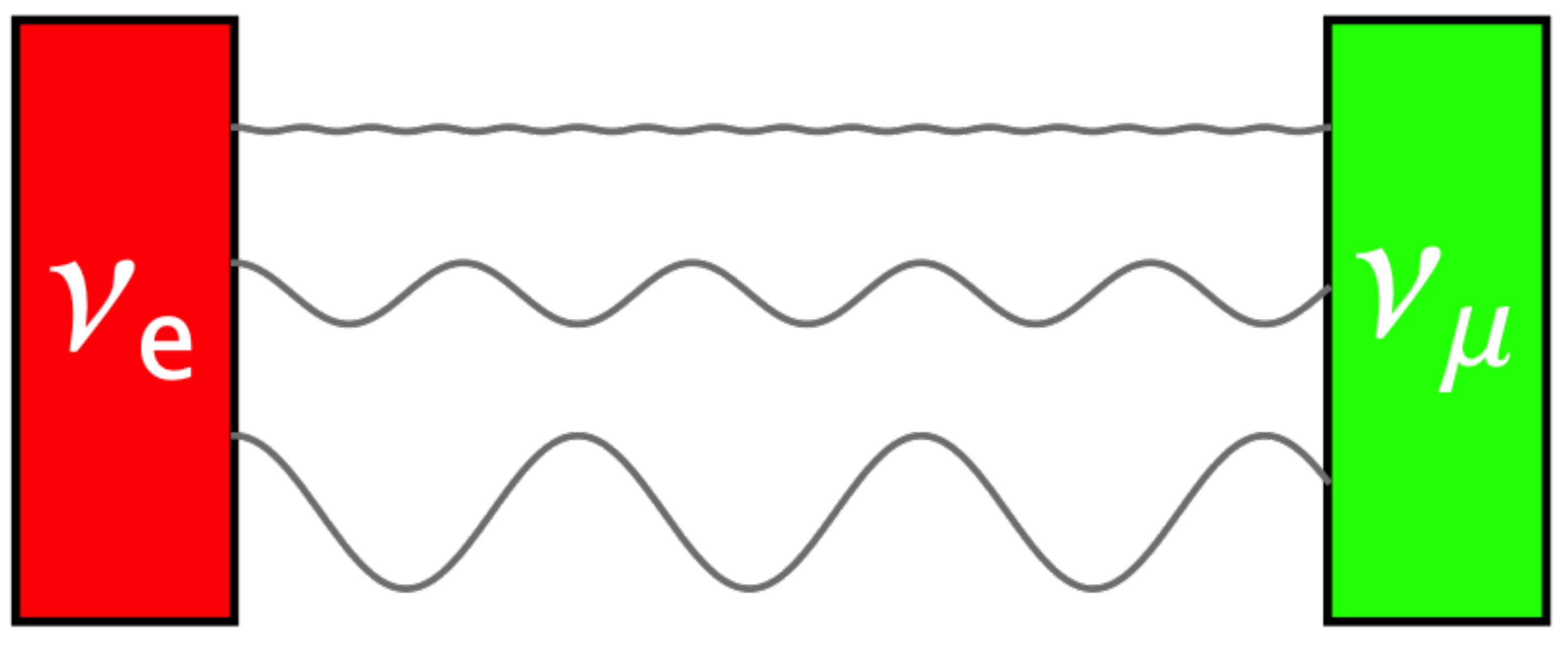}}
\caption{\small\sf\em  Schematic description of spontaneous neutrino transformations - i.e., neutrino oscillations.
The originally produced neutrino of electronic type - denoted as $\nu_e$ - can be thought of as a superposition of particles with given masses $m_i$, each evolving with different frequency $f_i=E_i/h$ where $E_i=\sqrt{(p c)^2 + (m_i c^2)^2}$. Therefore, there is a finite probability that, after propagation, a muon-type neutrino - denoted by $\nu_\mu$ - will be detected.}\label{fig2}
\end{figure}

From the late 1960s until the end of the last century, solar neutrino experiments began taking data,  finding results that {\em constantly disagreed} with expectations by a factor of the order 2-3.
But neither experiments nor astrophysical expectations were to blame.
Today we attribute this circumstance 
to an important characteristic of these curious particles, which we  understand well and have verified in great detail.
We refer to the phenomenon that was originally called neutrino `transition' or `transmutation' and nowadays it is commonly  
called `neutrino oscillations'. 
This is confirmed by experiments, it is precisely measured, and allows us to interpret the data correctly. 
(For a pedagogical introduction, see~\cite{ooo}; for a schematic description, see~Figure~\ref{fig2}.)

Here we would like to limit ourselves to examining the way the idea was formulated, which once again emphasises the importance of multidisciplinarity.

It all starts with the discussion of certain particles with strong interactions, which apparently have nothing to do with neutrinos.
More precisely, the discussion concerned the neutral mesons $K^0$ and $\bar{K}^0$ endowed with `strangeness'
(particles that, currently, we think of  as bound systems containing the valence quarks $d\bar{s}$ and $s\bar{d}$ respectively). 
It was  Gell-Mann, Pais and Piccioni \cite{gp}  who suggested the idea that, since strangeness was violated in their spontaneous decays, it was possible to consider 
a mixing between the two mesons, which lifted the exact degeneracy between the particle and antiparticle masses. Therefore,
the transition amplitude between the two meson states is non-zero, and interesting transformations are expected, as verified experimentally in 1964. 

This proposal was immediately considered very stimulating:\\
$\bullet$ Bruno Pontecorvo was the first to extend it to neutrinos (1957) \cite{bp}, although the specific version he proposed, that of a transition between neutrinos and antineutrinos, is not the currently accepted one. But it was again he who overcame these limitations, providing ten years later the {\em formal description of neutrino transitions} propagating in vacuum, which is now accepted  \cite{bp2}.\\
$\bullet$ The idea of a conversion between  {\em different} neutrinos was proposed in 1962
by Shoichi Sakata  and coauthors.
%$\nu_e$  associated with the electron and $\nu_\mu$ associated with the muon. 
Even if the mathematical description of the  transition was still missing, 
the description of the mixing  between neutrinos in  \cite{nagoya} agrees  with the modern  one.\footnote{Note in passing that the mixing of hadrons and that of leptons were originally motivated by hadron models \cite{gl,nagoya} that have now been superseded by the quark model.}\\
 Today, the mixing matrix between leptons is usually referred to as the `Pontecorvo-Maki-Nakagawa-Sakata matrix' to recall the valuable contribution of these pioneers,
even though, as we mentioned above, none of the original formulations were, strictly speaking, correct.  This is not surprising, because good ideas are usually stimulating and fruitful from the moment they appear, even when they are not fully defined, or even faulty in some detail.

A final interesting note in passing. The observation 
of another phenomenon  concerning the $K^0$ and $\bar{K}^0$ mesons - called `$K$-mesons regeneration' - 
inspired further theoretical advances concerning neutrinos. I am referring to the so-called `matter effect' or `MSW effect',
named after the proponents Mikheyev, Smirnov and Wolfenstein~\cite{msw}.
This effect implies that when neutrinos propagate in ordinary matter, their 
transformations change, provided certain conditions are met. To assess the significance of this effect, it is sufficient to evaluate, in the cases of interest, the following ratio 
\begin{equation}\label{emsw}
\varepsilon_{\mbox{\tiny MSW}}=
\left( \frac{E_\nu}{5\, \mbox{MeV}}\right)
\left( \frac{7.7\times 10^{-5}\, {\mbox{eV}^2}}{\Delta m^2\, c^4}\right)
\left( \frac{\rho\, Y_e }{100\, \mbox{g/cm}^3}\right)
\end{equation}
where $E_\nu$ is the energy of the propagating neutrinos, 
$\Delta m^2$ is the mass difference between the neutrinos, $c$ the speed of light, 
 $\rho$ is the density of matter where the neutrinos propagate, $Y_e$ the fraction of protons and hence electrons.
The matter effect is negligible when this ratio is small, but could be significant when it is of the order of unity or greater.
See again~\cite{ooo} for further discussion. 

I have not told the story of the reluctance to seriously consider these ideas, not because it lacks interest, but only because it would be too long.
In any case, today it is believed that the existence of `neutrino oscillations' has been conclusively demonstrated, and that the matter effect has also been definitively observed with solar neutrinos.
Recall that the 2015 Nobel Prize in Physics concerns precisely the observational verification of these phenomena.
Accurate (few \%) measurements  of the relevant neutrino parameters have been obtained, see e.g.~\cite{lisi}. 

\section{The beginning of a new science}\label{sez4}

As the Earth contains radioactive elements, the idea of reasoning about neutrinos produced in their decay, 
possibly estimating the amount of these elements and our ideas about the Earth's composition, 
seems  natural and reasonable. Today,  these are conventionally referred to as {\bf geo-neutrinos}.
 Note that neutron-rich radioactive elements 
 provide as a rule  electron antineutrinos.
 
Their first description  is by Eder in 1965 \cite{eder}, who pointed out the interest to observe the decay of 
$^{232}$Th e $^{238}$U above the minimum energy 
$E_\nu> [(m_n+m_e)^2-m_p^2]/(2 m_p)= 1.806$ MeV  in the reaction of Eq.~\ref{ibd}; 
however, his estimate of the flux is far from accurate.

A nice and simple argument from~\cite{gianni}  gives us a much better estimate:\\
$\bullet$ First, we note that, for each antineutrino produced by $^{232}$Th and $^{238}$U,  
about $\langle Q\rangle \sim 10$ MeV are released. \\
$\bullet$ We then assume that a part of the Earth's heat $H_\odot = 47\pm 2$ TW \cite{dd}
is due to radioactive decays - let us say about half, $H_{\mbox{\tiny rad}} \sim 20$ TW.\\
$\bullet$ At this point, it is easy to estimate the emission rate of the antineutrinos $\dot{N}_{\bar\nu_e}= H_{\mbox{\tiny rad}}/\langle Q\rangle \sim 10^{25}$ Hz 
and the corresponding flux
\begin{equation}
\varphi  \sim \frac{\dot{N}_{\bar\nu_e}}{4 \pi R^2} \sim \mbox{ few} \times 10^{6} /(\mbox{cm$^2$ s})
\end{equation}
where $R$ is the Earth's radius, the typical distance from which we expect the signal.

 I refer to~\cite{gianni,livia} and to the contributions of colleagues for a more complete introduction to geo-neutrinos, especially as regards the relevant 
 geophysical aspects, which I will limit myself to touching upon where appropriate, but will not elaborate in any detail. 

 \subsection{On the accuracy of particle physics expectations}
 The review just mentioned \cite{gianni}   is also clarifying with regard to geo-neutrino oscillations; it is noted there that they can be treated quite simply.\\
%The review just mentioned \cite{gianni}  is valuable also for what regards neutrino oscillations; it is observed there that their treatment 
 First, owing to the relatively small 
antineutrino energies of interest, $E_\nu<3.26$ MeV, the MSW effect is not large, and one can use 
the  oscillation probability for electron antineutrinos in vacuum $P_{\bar\nu_e\to \bar\nu_e}^{\mbox{\tiny (v)}}$. \\
Second, in view of the relatively large distances 
that antineutrinos travel, this reduces to a constant factor 
$P_{ee}^{\mbox{\tiny (v)}}=\langle P_{\bar\nu_e\to \bar\nu_e}^{\mbox{\tiny (v)}} \rangle$, given by the 
averaged on the distance of propagation:  %(which is sufficient to treat the contribution of the mantle):
$P_{ee}^{\mbox{\tiny (v)}}=(1-s_{13}^2)^2 \left[ \, 1- 2 s_{12}^2 (1- s_{12}^2) \, \right] + s_{13}^4$, 
where the notations for the mixing angles are as in \cite{lisi}. With the newest values and their 1$\sigma$ uncertainty ranges, 
given in \cite{lisi}, namely, 
$s_{12}^2=0.303\times (1\pm 1.1\%) $ and $s_{13}^2=0.0223\times (1\pm 3.1\%) $, we find the central value and its
1$\sigma$ uncertainty range (mostly due to  $s_{12}^2$) 
\begin{equation}
P_{ee}^{\mbox{\tiny (v)}}=0.553\times ( 1\pm 0.5\%)
\end{equation}
The error is small; thus, for the analysis of small datasets, $P_{ee}^{\mbox{\tiny (v)}}$ can be considered known. In this manner, the 
effect of oscillations factorizes and the formulae simplify.\\
We can note with Eligio Lisi
that the propagation effect in matter, which increases the probability a little, is of a similar size to the error mentioned above. to the error mentioned 
above.\footnote{The formula of propagation in constant density
 can be expanded linearly as:  $P_{ee}=0.553\times ( 1+ 0.58\, \varepsilon_{\mbox{\tiny MSW}} )$. Using 
  Eq.~\ref{emsw}, considering $\rho\, Y_e\sim 2$ g/cm$^3$ as in the mantle  
  and  considering   the observable region, we conclude that the matter effect increases the probability $P_{ee}$ by  
  0.4\% for  $E_\nu=1.8$ MeV and of twice this value for $E_\nu=3.3$ MeV.}
This is interesting, but not very important for the current data set.

It should be noted that the reaction in Eq.~\ref{ibd} remains the most important for the observation of geo-neutrinos. because of its relatively high probability.
%Note that the reaction of Eq.~\ref{ibd} remains the most important one for observing geo-neutrinos, owing to its comparably large probability.
Its modern determinations have rendered the uncertainty completely negligible, also thanks to detailed comparisons with the precise measurements of the average life of 
the neutron $\tau_n$. The most recent and accurate calculation, which includes this test, is in~\cite{na}. 

In short, we do not expect any significant uncertainty from the side of particle physics. 

 \subsection{Dependence on the number and distribution of radioactive elements}

Predictions of geo-neutrinos 
then make it possible to study their origins. To be precise, they depend\\
- on the quantities of radioactive elements\\
- on their arrangement in the Earth's interior.\\

The first factor can be estimated with models which combine geophysical, geochemical and cosmochemical arguments.
Although the total amount of elements presents uncertainties,  
the ratio 
\begin{equation}\label{ratto}
\frac{\mbox{Th}}{\mbox{U}}=3.9 
\end{equation}
from meteorites 
\cite{sun} seems more reliable.
The second factor relates to the fact that predictions change depending on where the elements are located.

 \begin{figure}[t!]
\centerline{\includegraphics[width=0.95\textwidth]{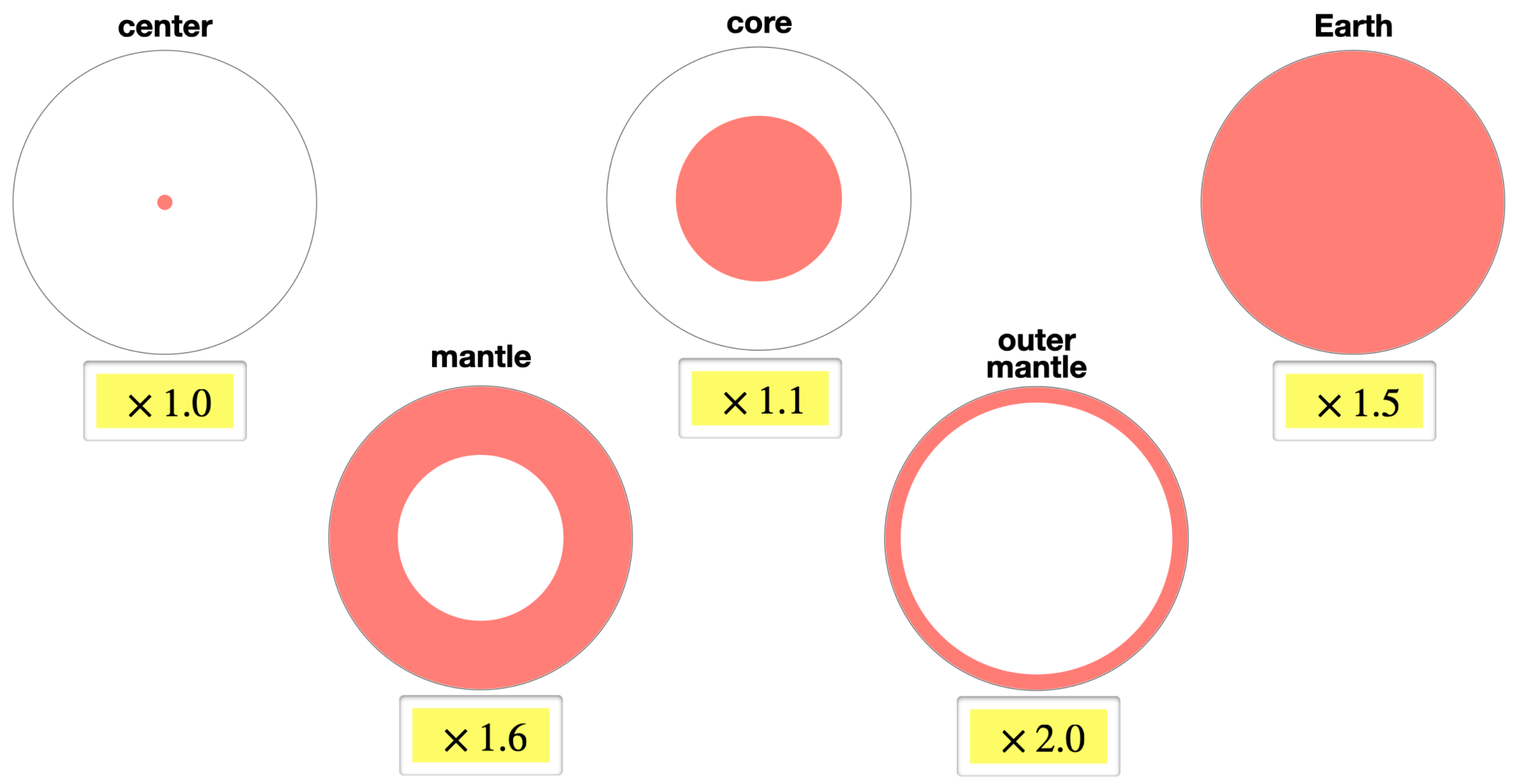}}
\caption{\small\sf\em  The 5 circles symbolise the structures of the Earth beneath the crust. It is assumed that 
the same amount of a certain radioactive species is uniformly distributed in various internal regions (red). 
The yellow panels indicate the corresponding values of the geometric factor $g$ defined in the text.
The first case (upper left, where the radioactive elements are in the center of the Earth) is the reference one.}\label{fig3}
\end{figure}

To clarify the point, and basing our discussion again on \cite{gianni} (their appendix), 
let us suppose that there are $N$ nuclei of $^{238}$U inside the Earth, whose lifetime is $\tau\approx 6.4$ Gyr.
 The component of the electron antineutrino flux  from one of the six radioactive species $i$ in the decay chain of   $^{238}$U 
 in a certain site is given by\footnote{Each species of the decay chain is assumed to be in secular equilibrium. The second longest lifetime in the $^{238}$U decay chain is that of $^{234}$U, which lasts 0.35 Myr; that of the $^{232}$Th decay chain is $^{228}$Ra, which lasts only 8 years. Therefore, the approximation should be quite reliable.}
\begin{equation}\label{fermil}
\frac{d \varphi_{i}}{dE_\nu} =g\times P_{ee}
 \times \frac{N/\tau}{4\pi R^2} \times BR_i\times \xi_i(E_\nu)
\end{equation}
where $i$ is the selected radioactive species~$i$,  
$\varphi_i$ is the corresponding electron antineutrino flux,  
 $BR_i$ is the branching ratio,\footnote{One includes the branching factor $BR_i< 1$
 when, for the selected radioactive species,  the decay chain bifurcates - i.e., for $^{232\mbox{\tiny m}}$Pa in the decay chain of   $^{238}$U and for $^{212}$Bi  in the decay chain of   $^{232}$Th; otherwise $BR_i=1$.}  
 $\xi_i$ is the normalized spectrum of the species (such that $\int_0^\infty  \xi_i(E_\nu)\ dE_\nu=1$), 
$R$ is the radius of the Earth and $g$ is a geometrical factor, that can be calculated easily as discussed in the appendix of the present writeup, 
supposing that the distribution of $^{238}$U   is approximatively spherically symmetric. This does not apply to the contribution of the crust, but presumably applies to the contribution of the mantle and of other internal structures, that is much more uncertain and that we are interested to discuss.

The value of $g$ for some representative cases is shown in Figure~\ref{fig3}. We observe that:\\
$\bullet$~the effect is large, presumably $g\sim 1.5$, and 
up to a factor of two; \\
$\bullet$~that the closer the radioactive elements are to the detector, the greater the flux; \\
$\bullet$~that a theoretically interesting case, when the radioactive elements are mainly (or only) in the mantle, leads to a value of the factor $g=1.6$ almost identical to the case when the element is uniformly distributed throughout the Earth  $g=1.5$.

This gives us an idea of the variations we could expect in the predictions. 
The same considerations can be repeated for $^{232}$Th, using  a different $N$ and $\tau\approx 20$ Gyr.   
% that describes the distribution of the nuclei, and $P_{ee}$
%describes antineutrino oscillations. 

 \subsection{First observations and results}
About twenty years ago, the KamLAND (Japan) detector provided the clues to the first observation of a signal   \cite{k1}; a few years later, this signal was confirmed by Borexino (Italy)~\cite{b1}. 

Both are ultra-pure scintillators, located in underground laboratories to minimise disturbances caused by interactions with cosmic rays. KamLAND is about three times larger than Borexino, and located at a site where the presence of nuclear reactors, which also produce electrons antineutrinos, causes more spurious events~\cite{recc}. However, the Fukushima accident (2011) led to the shutdown of many reactors and thus to better conditions for geoneutrinos research. 

As we discussed in the workshop, both detectors have progressed - see e.g.~\cite{k2,b2,k3} - and provided further valuable measurements of geo-neutrinos fluxes. 
Signal detection is out of the question; both uranium and thorium contribute to this, in a proportion not precisely probed, but not inconsistent with Eq.~\ref{ratto}. 
Moreover, the data - especially those from the KamLAND - 
point to the fact that the Earth's heat is not entirely due to radioactive disintegrations.

The flux of `local' geoneutrinos, associated with the continental crust, can be reliably estimated (it should be larger than 30\% in KamLAND) and can be subtracted from the 
observed flux,  to obtain the contribution of the mantle; this, assuming spherical symmetry, should be the same in KamLAND and Borexino \cite{lisio}.

The two results are not mutually incompatible and suggest that we could have an evidence for the presence of geoneutrinos from the mantle - mainly from the Borexino data~\cite{b2}. however, the agreement between the two experiments on this point is not excellent.\footnote{Let us summarize what we have learned  on the signal from the mantle, adopting  TNU units \cite{mant04}. Ref.~\cite{gianpi} 
used $S_{M}^{\mbox{\tiny BX}}=21.2_{-9.0}^{+9.6}$ TNU \cite{b2} and $S_{M}^{\mbox{\tiny KL}}=6.0_{-5.7}^{+5.6}$ TNU \cite{wat}
(which do not agree perfectly) and  estimated from them the combined value  $S_{M}^{\mbox{\tiny BX+KL}}=8.9_{-5.5}^{+5.1}$ TNU.  
More recent results by KamLAND~\cite{k3} seem to suggest a slightly lower mantle flux.}

%leading roughly to $S_{M}^{\mbox{\tiny KL}}\approx 4\pm 5$ TNU (our estimation)

%$S_{M}^{\mbox{\tiny BX+KL}}=8.9_{-5.5}^{+5.1}$ TNU, i.e.,  $\varphi\approx (1\pm 0.6)\times 10^6$/(cm$^2$s) for the reference Th/U=3.9 ratio.

It would be very interesting   if precise measurements revealed that the fluxes of geoneutrinos from the mantle are indeed site-dependent; or in other words, if we were to admit that the distribution of radiogenic elements deviates significantly from spherical symmetry, as is the case, e.g., with the flux of heat on the Earth's surface.

%
%%Let us now try to see these results in perspective, cultivating the hope of future, more precise measurements.
%Subtracting the contributions of ``local'' geo-neutrinos associated with the continental crust (which it is expected to be 30\% larger in KamLAND)
%we expect the remaining signal to be the mantle contribution, which assuming spherical symmetry, should be the same.
%The two outcomes are not mutually incompatible and suggest~\cite{gianpi} that we may see a first hint of mantle geoneutrinos (especially the Borexino data). However, it would be even more exciting  if more precise measurements would indicate that geo-neutrino fluxes from the mantle are indeed site-dependent; or in other words, to realise that the distribution of radiogenic elements deviates significantly from spherical symmetry.
%

%
%\footnote{Ref.~\cite{gianpi} 
%combines $\varphi_{M}^{\mbox{\tiny BX}}=21.2_{-9.0}^{+9.6}$ TNU \cite{b2}
%and   $\varphi_{M}^{\mbox{\tiny KL}}=6.0_{-5.7}^{+5.6}$ TNU \cite{wat}, finding 
%$\varphi_{M}^{\mbox{\tiny KL}}=8.9_{-5.1}^{+5.5}$ TNU, i.e.,  $\approx (1\pm 0.6)\times 10^6$/(cm$^2$s).
%The agreement is moderate/marginal and  it does not improve after~\cite{k3}, which  
%roughly suggests a still smaller flux $\varphi_{M}^{\mbox{\tiny KL}}\approx 4.5\pm 6$ TNU.}
%%corresponding to $\varphi_{M}^{\mbox{\tiny KL}}=0.67_{-0.64}^{+0.63}\times 10^6/($cm$^2$s).
%%$\phi_{M}^{\mbox{\tiny KL}}\approx (0.5\pm 0.6) \times 10^6/($cm$^2$s).  
%

It is undeniable that the KamLAND and Borexino observations, although still small in statistical terms and significance, have well and truly inaugurated the science of geo-neutrinos and 
provided the first quantitative indications, useful to geophysicists. I refer to the conference proceedings for a detailed and in-depth discussion of these aspects.

\section{Discussion and dedication}\label{sez5}

The beginning of geo-neutrino science is an awaited and welcome event, which follows the many and beautiful results in astrophysics obtained by means of neutrinos;
for an overview of the current situation, see e.g.~\cite{e}.
There is no shortage of things still to be done; in addition to the objectives just recalled (such as probing the Th/U ratio, mantle contribution, etc.) there are many other interesting ones, e.g.~the search for
new and unexpected geo-neutrino sources, hypothetical site dependencies, observation of the direction of arrival of anti-neutrinos, or attempts to detect geo-neutrinos from $^{40}$K decay.

 The near future of the field is related to the performance of the SNO+ and JUNO scintillation detectors, the former already operational, with a mass of 0.78 kton and not too close to nuclear reactors~\cite{sno+}, and the latter which (while operating close to nuclear reactors) will be 20 times larger than KamLAND and will soon become operational~\cite{juno}.
 The more distant future is linked to new projects, which could for example involve new detectors operating far from the continental crust, and thus more directly sensitive to the contribution of the Earth's mantle.
  
  However, the interest shown by the scientific community in such a promising field of research does not really seem to live up to its importance.
Based on past experience with solar neutrinos, one could attribute this unfortunate circumstance to the multidisciplinary nature of neutrino science - see \cite{jb} for illuminating remarks in this regard. 
This is why I wanted to start with a broad excursus, to argue how the long lines of the history of science do not always coincide with those dictated by  current urgencies or trends.

I would like to make an attempt to rephrase the point even more directly. To say that ``good science takes geological time" in this meeting may be considered a joke, but it usually takes a long time to get there, especially in a relatively mature field of research. For this reason, it is rare to see the birth of a new discipline as is happening to us with geoneutrinos. I believe this puts us in a position of special responsibility for the future development of a science that offers us the opportunity to better understand the interior of our planet. For progress to occur, I believe it is helpful to recognise the deep historical roots and even more so the links between this and other vibrant disciplines.

\bigskip
With these considerations in mind, I would like to dedicate this essay  to the memory of my colleague and friend Giovanni ``Gianni'' Fiorentini, who passed away in June 2022. He gave numerous
and valuable contributions to science. His theoretical studies on geo-neutrinos, in particular, have had a decisive character for the progress achieved, in particular those of the Borexino experiment.
I bring to all the greetings of three scientists, close collaborators of him: Marcello Lissia (Cagliari), Fabio Mantovani (Ferrara), Francesco Villante (L'Aquila). Before coming here, I asked them to tell me more about him and I was struck, though not surprised, by an almost identical sentence they all said\begin{quote}
{\em\small Gianni is the best physicist I have ever worked with.} 
\end{quote}
We will miss you very much, Gianni.

\paragraph*{Acknowledgments}

I am grateful to Gianpaolo Bellini and Lino Miramonti for the invitation, to Fabio Mantovani and Eligio Lisi for valuable discussions, and to all the participants of the meeting for fine reports and useful exchanges of ideas. Work carried out with partial support from the Italian Research Grant number 2017W4HA7S ``NAT-NET: Neutrino and Astroparticle Theory Network,'' under the PRIN 2017 program funded by MIUR.

\newpage
 \appendix
 %\section{Hydrogen and helium}

\section{My notes for the meeting}\label{summ}
On October 27, 2022, the Italian Physical Society (SIF) organized  in Milan a study day, in which a {\em new way of looking at the interior of our planet} was discussed. Resorting to a medical metaphor, we can say that now, in addition to the usual 
diagnostic tools (thermometer, auscultation, ultrasound) we have learned to take 
x-rays.
Indeed, 
in addition to measurements of emitted heat, local gravity, and magnetic field, 
to the study of earthquakes and their propagation, 
we also know how to detect certain very penetrating particles 
that come from the Earth's natural radioactivity and are called geoneutrinos. And, as we shall see, this does not exhaust the novelties.

The very interesting study day was prepared and coordinated by {\em Angela Bracco, Gianpaolo Bellini and Lino Miramonti}
and divided into four sessions.

1) The first presented the {\bf state of knowledge} of the planet's interior. 
{\em Cinzia Farnetani} showed that the actual compositions of the Earth's mantle and crust critically depend on factors such as.
$(i)$~the initial composition of the Earth, which is similar but
not identical to any known meteorite; 
$(ii)$~the geodynamic processes that animate the mantle, such as convection stirring rocks 
and partial melting
producing continental crust, in which the concentration of radioactive elements 
{heat-producing elements} is high.

 {\em Chris Davies} discussed in detail expectations about the nature of Earth's heat, 
 resorting to complex theoretical modeling and reasoning about structures present at the base of the mantle and in the outer core, of which we have observational clues.
The number and quality of problems debated are remarkable,
and the investigations of structures such as the ``plumes'' (plumes) of hot matter in the mantle
involving deviations from the spherical distribution.

2) 
The second session illustrated {\bf the use of new particles} for geophysical and geochemical investigations and beyond.
{\em Jacques Marteau} showed images obtained 
with the technique of muon-based tomography (muography); specifically, 
the interior of volcanic calderas, which makes it possible to follow the evolution of eruptions.
{\em Nicola Rossi} argued %very convincingly. 
the possibility of even investigating stellar interiors 
by means of neutrinos, presenting the admirable results of the experiment {\sc Borexino} (Italy).  In particular, he presented unequivocal observations 
of the effects of the eccentricity of the Earth's orbit on the number of neutrinos observed during the year: in other words, {\sc Borexino} measured the 
small difference between winter and summer observations, which consists of a few percentage parts.

3) 
The third session was the central session of the conference. In fact, the following were discussed 
the first observations of the so-called {\bf geoneutrinos} 
that is, antineutrinos produced spontaneously 
by the disintegrations of radioactive elements contained within the Earth.  
These were the results of {\sc KamLAND} (Japan),
presented by {em Nanami Kawada} and those of {sc Borexino}, illustrated by {\em Livia Ludhova}. 
The data collected for the past few years 
concern the decay chains of uranium-238 and thorium-232, which we are close to identifying individually;
the geoneutrinos produced by potassium-40 do not have sufficient energy to be revealed. Integrating the 
helpful remarks of {\em Fabio Mantovani}, who could not be present in person, the two speakers  
summarized the framework of expectations, 
discussing what information can be obtained about the chemical composition of our planet, and comparing it with observations.
The existing data do not contradict, indeed corroborate the most accepted theories and are mutually consistent, although they consist of
at the moment of relatively small samples.    
These advances indicate that we have begun to probe the mechanisms of planetary formation, those involving the 
distribution of elements, the amount of radioactive material and its contribution
to the heat radiated from the Earth, etc.: considerations pertaining to diverse and challenging fields.

4) The last session gathered elements for an {\bf assessment of the field of research} that opened up with this type of observations. 
{\em Jo\~ao Coelho} reminded of the existence of other neutrinos of natural origin and with higher energies: $(i)$~atmospheric neutrinos, which thanks to the refraction effect on terrestrial matter (MSW effect) 
could provide us with definitive validation of the way the neutrino mass spectrum is structured, and perhaps teach us something about the mantle, 
if we have a sufficiently large detector; $(ii)$~those of cosmic origin, which--as shown by the {\sc IceCube} (South Pole) experiment--undergo partial attenuation (=absorption) in crossing our planet.  The concluding paper, by yours truly,  
pointed out the many scientific links between investigations of the Earth's interior and those of the Sun,
and argued how different assumptions about the distribution of radioactive elements result in a 25\% difference in the 
number of geoneutrinos in the most plausible cases or even 100\% in the most extreme ones.

Extensive discussions accompanied and enriched the reports highlighting a vital field of research with great prospects for development.
The consensus emerged that there will be {\bf further progress} with future measurements from the experiments {\sc SNO+} (Canada),  
{\sc JUNO} and {\sc Jinping} (China); it would also be forward-looking to
consider new ones that relate more directly to the contribution of geoneutrinos produced in the mantle, to be carried out at some site away from the continental crust.

 \section{Geometrical factor}

Let us consider  the decay of $^{238}$U in the Earth, 
and denote the corresponding rate of disintegrations
by 
$$d\dot{N}=\dot{n}(\vec{r}) \, d\vec{r}=  \frac{ n(\vec{r}) d\vec{r}}{\tau} %\mbox{ at point }\vec{r}
$$ 
where 
 $\vec{r}$ is a vector from the center of the Earth to the point of interest,
 $d\vec{r}$ the infinitesimal volume, 
  $n$ the density is the radioactive species and  
$\tau$ its lifetime. % of this species.
The flux a certain type of neutrinos, for an observer on the surface of the Earth, 
which is produced by a certain species $i$ produced in 
 the decay chain of $^{238}$U, 
is given by\footnote{This agrees with Eq.~\ref{fermil} after integrating on the possible energies $E_\nu$, noting that, for the sake of brevity, we have omitted two constant factors:  the oscillation factor $P_{ee}$ and the branching ratio $BR_i$.}
$$
\varphi_i=\int_{S_\oplus} \frac{d\dot{N}}{4\pi\, d^2} \quad
\mbox{ with }\quad
 \left\{ 
\begin{array}{c}
d(\vec{r})= | \vec{R}-\vec{r}|^2\\[1ex]
S_\oplus=\{\vec{r}\mbox{ such that }r\le R \}
\end{array}
\right.
$$
where $\vec{R}$ is the vector from the observer to the center of the Earth, $R=|\vec{R}|$ is the Earth's radius; thus 
 the integral is taken over the whole Earth.

Assuming an approximate spherical symmetry,
which should be sufficiently accurate to discuss the contribution of the mantle - not the continental crust -  
 the density $n(\vec{r})$ is a function only of the distance from Earth's center 
$$r=|\vec{r}|$$
and we can carry out the integrals over the angle, 
by finding
$$
\varphi_i=\frac{R}{2} \int_0^1  du \ u \ \log\frac{1+u}{1-u}  \ \dot{n}(u)
$$
where 
$$
u=\frac{r}{R}
$$
Next, let us now compare the effect on the flux from different geometrical configurations.  

$\bullet$ 
Assuming that the $N$  radioactive nuclei (a number)  are distributed in a {\em shell} of thickness $dr$,  centered, we have, 
$$
\dot{n}=  \frac{N}{4 \pi r^2\ dr\  \tau}
$$
This gives immediately,
$$
\varphi_i=\frac{1}{2 R} \   {dr}\ r \ \log\frac{1+u}{1-u} \    \frac{N}{4 \pi r^2\ dr\  \tau} = \frac{N/\tau}{4\pi R^2 }\times  \frac{1}{2 u} \ \log\frac{1+u}{1-u} 
$$
The function in the last factor grows with $u=r/R$; thus the closer the shell is to the observer, the more relevant it is. 
(Compare with Newton's theorem, that every spherical shell gives the same force; note that in that case we speak of 
of a {\em vector quantity} while here we are dealing with  {\em scalar quantities} - the density, the flux).

$\bullet$  Let us then consider the case in which the elements are only inside a {\em sphere} of radius $r$, centered, 
and are uniformly distributed due to the original distribution and mixing mechanisms within it. 
When we denote by $N$ the total number of radioactive nuclei, as a above,  we have
$$
\dot{n}=\frac{N}{\frac{4 \pi}{3} r^3\ \tau}= \frac{3 N}{4 \pi r^3\ \tau}
$$
Evaluating the integral over the radius, we find 
that the flux is given by 
$$
\varphi_i=\frac{ \dot{n}}{2R}  \   f(u)  
$$
%where
%$$
% u=\frac{r}{R}
%$$
where\footnote{For verification purposes, let's consider the limit 
$f(\epsilon)=\frac{2}{3} \epsilon^3 + O( \epsilon^5)$.
If the sphere is small (near the center of the Earth)
the following applies
$
\varphi_i= \frac{r^3\ \dot{n} }{3} \frac{1}{R^2} =   \frac{N/\tau}{4\pi R^2} 
$
which is the obvious and expected result.}  
$$
f(u)=u- \frac{1-u^2}{2}  \ \log\frac{1+u}{1-u} 
$$
We conclude that the radiation from the sphere can be written
$$
\varphi_i=\frac{N/\tau}{4\pi R^2} \times \frac{3}{2} \frac{f(u)}{u^3}
$$

$\bullet$ Similarly, the flux 
from a shell of finite dimension, enclosed among $r_1<r<r_2$,  
is given by,
$$
\varphi_i=\frac{N/\tau}{4\pi R^2} \times  
  \frac{3}{2} \frac{f(u_2) - f(u_1)}{u^3_2- u_1^3}  \mbox{ where } u_i=\frac{r_i}{R}
$$

 \begin{table}[t]
 \centerline{
 \begin{tabular}{l|ccccc}
	& average crust & upper mantle & lower mantle & outer core & inner core\\
& (35 km) &  (635 km) &  (2221 km) &  (2259 km) &  (1221 km) \\ \hline
fraction of mass &	0.5\% & 	17.8\% & \bf	49.2\% & 	\bf 30.8\% & 	1.7\% \\ 
 fraction of 
volume &	1.6\% & 	26.7\% & 	\bf 55.4\% & 	15.6\% & 	0.7\% \\ 
 annuli' 
fraction & 	1.1\% & 	18.8\% & \bf 	\bf 50.2\% & \bf	26.2\% & 	3.7\% \\ 
weight as 
1/$d^2$  &	3.8\% & 	\bf 35.5\% & 	\bf 49.1\% & 	11.1\% & 	3.8\%  \\ 
solid angle from ground	& 10.5\% & 	\bf 34.2\% & 	\bf 39.1\% & 	14.4\% & 	1.9\% 
\end{tabular}}
\caption{\em\small Various structures in the Earth and their comparison
(note that we introduce a concept of `average crust' only to give an initial tentative estimate, 
but this assumption is known that the crust deviates from spherical symmetry).
Fractions larger than 1/4 are emphasized. 
\label{trb}}
\end{table}

The above results can be presented in terms of a geometrical factor $g\ge 1$, defined as 
$$
\varphi_i= g\times \varphi_0 \mbox{ where } \varphi_0=\frac{N/\tau}{4\pi R^2}
$$
where $\varphi_0$ is the naive estimation of the flux, corresponding to the fictitious, reference  
case, when the radioactive particles are in the center of the Earth and the geometrical factor is minimal, 
 $g=1$. From the positions of the structures in the Earth (radii of inner and outer mantle, of the core, etc) 
 the results of Figure~\ref{fig3} follows; more remarks in Table~\ref{trb}.

\tableofcontents
\end{document}